\documentclass[
  aps,
  prb,
  twoside,
  twocolumn,
  showpacs,
  floatfix,
  superscriptaddress,
  10pt,
  preprintnumbers,
  citeautoscript,
]{revtex4-1}

\usepackage{amsmath}
\usepackage{amssymb}
\usepackage{booktabs}
\usepackage{comment}
\usepackage{float}
\usepackage{glossaries}
\usepackage{graphicx}
\usepackage{tabularx}
\usepackage{units}
\usepackage[dvipsnames]{xcolor}

\usepackage[colorlinks=false,hidelinks]{hyperref}
\hypersetup{
    pdftitle={Luminescence quenching via deep defect states: A recombination pathway via oxygen vacancies in Ce-doped YAG},
    pdfauthor={Christopher Linderälv, Daniel Åberg, and Paul Erhart},
    colorlinks=false,
}

\newcommand{\phys}{
  Chalmers University of Technology,
  Department of Physics,
  Gothenburg, Sweden
}
\newcommand{\llnl}{
  Physical and Life Sciences Directorate,
  Lawrence Livermore National Laboratory,
  Livermore, California, USA
}

\newcommand{\VO}{$\text{V}_{\text{O}}$}
\newcommand{\VOneu}{$\text{V}^{\times}_{\text{O}}$}
\newcommand{\VOpos}{$\text{V}^{\bullet}_{\text{O}}$}
\newcommand{\VOdpos}{$\text{V}^{\bullet\bullet}_{\text{O}}$}
\newcommand{\ceyag}{YAG:Ce}
\newcommand{\Cepos}{$\text{Ce}_{\text{Y}}^{\bullet}$}
\newcommand{\Ceneu}{$\text{Ce}_{\text{Y}}^{\times}$}
\newcommand{\Ceex}{$\text{Ce}_{\text{Y}}^{\ast}$}

\renewcommand{\vec}[1]{\ensuremath\boldsymbol{#1}}
\renewcommand{\epsilon}{\varepsilon}

\newacronym{cbm}{CBM}{conduction band minimum}
\newacronym{cc}{CC}{configuration coordinate}
\newacronym{ctl}{CTL}{charge transition level}
\newacronym{dft}{DFT}{density functional theory}
\newacronym{hr}{HR}{Huang-Rhys}
\newacronym{LuAG}{LuAG}{Lu$_3$Al$_5$O$_{12}$}
\newacronym{pes}{PES}{potential energy surface}
\newacronym{si}{SI}{Supporting Information}
\newacronym{vbm}{VBM}{valence band maximum}
\newacronym{zpl}{ZPL}{zero-phonon line}

\AtBeginDocument{
\heavyrulewidth=.08em
\lightrulewidth=.05em
\cmidrulewidth=.03em
\belowrulesep=.65ex
\belowbottomsep=0pt
\aboverulesep=.4ex
\abovetopsep=0pt
\cmidrulesep=\doublerulesep
\cmidrulekern=.5em
\defaultaddspace=.5em
}

\begin{document}

\title[
  Luminescence quenching via deep defect states:
  A recombination pathway via oxygen vacancies in Ce-doped YAG
]{
  Luminescence quenching via deep defect states: \texorpdfstring{\\}{}
  A recombination pathway via oxygen vacancies in Ce-doped YAG
}

\author{Christopher Linder\"alv}
\affiliation{\phys}
\author{Daniel {\AA}berg}
\affiliation{\llnl}
\author{Paul Erhart}
\email{erhart@chalmers.se}
\affiliation{\phys}

\begin{abstract}
Luminescence quenching via non-radiative recombination channels limits the efficiency of optical materials such as phosphors and scintillators and therefore has implications for conversion efficiency and device lifetimes.
In materials such as Ce-doped yttrium aluminum garnet (YAG:Ce), quenching shows a strong dependence on both temperature and activator concentration, limiting the ability to fabricate high-intensity white-light emitting diodes with high operating temperatures.
Here, we reveal by means of first-principles calculations an efficient recombination mechanism in YAG:Ce that involves oxygen vacancies and gives rise to thermally activated concentration quenching.
We demonstrate that the key requirements for this mechanism to be active are localized states with strong electron-phonon coupling.
These conditions are commonly found for intrinsic defects such as anion vacancies in wide band-gap materials.
The present findings are therefore relevant to a broad class of optical materials and shine light on thermal quenching mechanisms in general.
\end{abstract}

\maketitle

\section{Introduction}

Optical materials with wide band gaps, such as phosphors and scintillators, enable the conversion of photons and high-energy radiation into one or several photons of lower energy.
This allows for applications in e.g., lighting (phosphors), radiation detection (scintillators) and lasing.
The performance of these materials is limited by non-radiative decay processes, most commonly via phonons and/or defects, which compete with the conversion into photons.
Luminescence becomes in particular quenched at higher temperatures, which imposes limits e.g., with regard to the maximum operating power or the operation temperature.
Despite extensive research, the mechanisms responsible for luminescence quenching are still not fully understood \cite{BacRonMei09, UedDorBos15}.

\begin{figure*}
    \centering
    \includegraphics[]{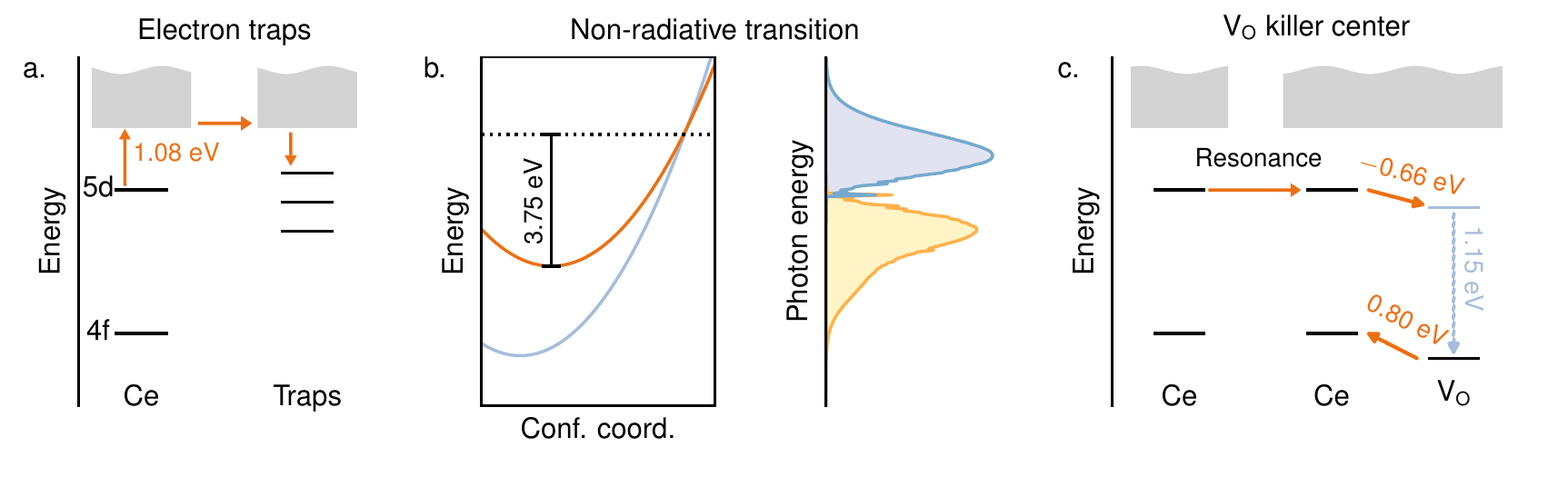}
    \caption{
        Schematic illustration of the mechanisms considered in the present study along with the highest energy barrier along each respective pathways.
        a) Capture of conduction band electrons at electron traps.
        b) Landscape cross-over by thermal excitation.
        c) Non-radiative recombination via oxygen vacancies.
    }
    \label{fig:mechanisms}
\end{figure*}

Cerium doped yttrium aluminum garnet (YAG:Ce, Y$_{3-x}$Ce$_{x}$Al$_5$O$_{12}$) is a yellow phosphor that is widely employed in rare earth-based solid state lightning \cite{NisTanFuj11, LinKarBet16}.
The optical transitions that are exploited in solid state lightning occur between $4f^15d$ and $4f^05d^1$ states of Ce.
Several mechanisms have been suggested to explain quenching in this material \cite{LinErhBet18, LinBetSha20} (\autoref{fig:mechanisms}):

(\emph{i})
In principle it could be possible for the system to undergo a \emph{thermally activated landscape crossover} via multiphonon processes between the ground (Ce:$4f^15d^0$) and excited states (Ce:$4f^05d^1$) \cite{BleBla79, Bla88}.
In this regard, it is has been observed that the Debye temperature increases with Ce concentration resulting in a larger high-frequency phonon population at lower temperatures, which in turn enhances multiphonon processes \cite{GeoPelDan13}.
Earlier studies, however, concluded that emission of high frequency phonons could not explain the temperature dependence of the non-radiative decay in YAG:Ce \cite{Web73}.

(\emph{ii}) While the onset temperature for luminescence quenching is above \unit[600]{K} for Ce concentrations below 1\%, it drops to around \unit[400]{K} if the Ce content reaches the percent level \cite{BacRonMei09}.
Based on this observation, the reduction in luminescence has been attributed to \emph{thermally activated concentration quenching}, in which the excitation energy is resonantly transferred between Ce atoms until it is eventually dissipated at a, not further specified, killer center.

(\emph{iii}) Finally, according to the \emph{thermal ionization model} an electron is emitted from the excited state of Ce to the conduction band edge ($4f^05d^1\rightarrow4f^05d^0+e'$) or a defect level.
The rate of this process has been shown to be much faster than the thermal crossing of landscapes rate in dilute Ce doped YAG at temperatures above \unit[573]{K} \cite{UedDorBos15}.

While a single mechanism can dominate under certain conditions, such as high Ce content or low temperatures, luminescence quenching is likely due to a combination of the above mentioned processes \cite{LinBetSha20}.
It is, however, difficult to discriminate their contributions based on experimental analysis alone.
This pertains in particular to mechanisms (\emph{ii}) and (\emph{iii}), which involve electron traps and/or killer centers provided by defects, the properties of which are notoriously difficult to access experimentally \cite{UedDorBos15, ZycBreGlo00}.

Here, using first-principles calculations, we analyze the mechanisms described above and identify a detailed recombination pathway involving oxygen vacancies.
While our calculations are specific for \ceyag{}, the key parameters of the proposed mechanism (deep transition levels and strong localized electron-phonon coupling) are generic and can be found in many other wide-band gap materials, including oxides \cite{LinLinErh18}.
Our results thereby point to specific mechanisms by which defects contribute to non-radiative recombination processes.
This insight is relevant for a broad class of materials and enables more directed approaches to mediating and controlling non-radiative decay processes.
As part of our analysis of luminescence quenching, we also probe the coupling between optical transitions and vibrational degrees of freedom using the generating function approach \cite{Mar59}.
We demonstrate that the phonon sidebands in Ce absorption and emission spectra, which have been previously associated with localized defect modes \cite{Rob79, BacRonMei09}, are in fact due to a superposition of many delocalized modes, an important finding in the context of the thermally activated landscape crossover model.

\section{Methodology}

\subsection{Lineshape}
\label{sect:methodology}

Optical lineshapes are modelled using the generating function approach \cite{KubToy55, Lax52, Mar59} under the Franck-Condon, parallel mode, and low temperature approximations, which results in a computationally feasible scheme \cite{AlkBucAws14}.
To begin with, we define the coordinate $Q_{\nu}$ as
\begin{align}
  Q_{\nu} &= \sum_{a}\sqrt{m_a}({\vec{R}^{i}_a} - {\vec{R}^f_a})\cdot {\vec{u}}_{\nu, a},
\end{align}
where $\vec{u_{\nu}}$ is the normalized displacement vector of phonon branch $\nu$.
The \emph{partial} \gls{hr} factor is defined as $S_{\nu} = \frac{1}{2} \omega_k Q_{\nu}^2$.
In the parallel mode approximation the normal modes of the initial and final states, and hence $Q_{\nu}$, are assumed to be related by a simple translation.

The lineshape function for absorption is
\begin{align}
  A(\omega_{eg} - \omega) &= \frac{1}{2\pi} \int_{-\infty}^{\infty} e^{i\omega t - \kappa t}G(t) \,dt,
\end{align}
where $\kappa$ determines the lifetime broadening of individual contributions,
$\omega_{eg}$ is the angular frequency corresponding to the transition between the ground and excited state, and the generating function $G(t)$ is given by
\begin{align}
 G(t) &= \exp\Big[S(t)-\sum_\nu S_{\nu} \Big].
 \label{eq:genFun}
\end{align}
Here, $S(t)$ is the Fourier transform of the spectral function
\begin{equation}
S(\omega) = \sum_{\nu} S_{\nu}\delta(\omega -\omega_{\nu}),
    \label{eq:spectral-function}
\end{equation}
where the $\delta$-functions in practice are approximated with normalized Gaussian functions.
The emission intensity is proportional to $\omega^3A(\omega)$ and the excitation intensity is proportional to $\omega A(\omega)$\cite{BaiBloBar13}.

\subsection{Computational details}
\label{sect:computational-details}

Electronic structure calculations were performed within the framework of \gls{dft} using the projector augmented wave method \cite{Blo94, KreJou99} as implemented in the Vienna ab-initio simulation package \cite{KreFur96, KreHaf93} with a plane wave cutoff energy of \unit[500]{eV}.
The exchange-correlation potential was approximated with the PBE functional \cite{PerBurErn96}.
Throughout this work, the DFT+$U$ method \cite{DudBotSav98} was used with $U=\unit[2.5]{eV}$ applied to Ce-4$f$ states, which has been shown to yield accurate positions of the Ce-4$f$ states in a range of phosphors \cite{CanChaBou11}.
The calculations on Ce defects in the excited state were performed by constraining the occupations of $4f$ and $5d$ states.
Spin-orbit coupling was included in the emission spectrum by the addition of an extra peak shifted by \unit[0.35]{eV} and redistribution of the spectral weight.
The shift was extracted from the energy difference of the $^2F_{5/2}$ and $^2F_{7/2}$ states in Ref.~\citenum{IvaOgiZyc2013}, while the intensity ratio is given by the multiplicities.

Most defect calculations were carried out using the 160-atom conventional (simple cubic) unit cell while the Brillouin zone was sampled using a zone centered $2\times 2\times 2$  $\boldsymbol{k}$-point mesh.
The ionic positions were allowed to relax until all forces fell below \unit[20]{meV/\AA}.
Phonon and lineshape calculations were performed using a 320-atom supercell and the Brillouin zone was sampled at the zone center only.
An extended Kr\"oger-Vink notation is used where
Ce$_{\text{Y}}^{\times}$ is a short notation for Ce-($4f^15d^0$),
Ce$_{\text{Y}}^{\ast}$ for Ce-($4f^05d^1$) and Ce$_{\text{Y}}^{\bullet}$ for Ce-($4f^05d^0$).

\subsection{Semi-local vs hybrid functionals}
\label{sect:hybrid-problem}

The PBE functional yields a band gap of  \unit[4.6]{eV}, which, as is common for semi-local functionals, is significantly smaller than experimental values that are found in the range from \unit[6.4]{eV}\cite{SlaOliChr69} to \unit[7.7]{eV} as discussed in Ref.~\citenum{NinJiDon16}.
The band gap underestimation can be compensated using hybrid functionals such as PBE0 \cite{AdaBar99}, which using a mixing parameter of 0.32 yields a band gap of \unit[7.8]{eV}, in much better agreement with experimental data.
When going from PBE to PBE0 \gls{vbm} and \gls{cbm} shift by \unit[2.1]{eV} and \unit[1.1]{eV}, respectively (see Fig.~S4).

For the oxygen vacancies, we therefore also carried out PBE0 calculations, for which we employed 80-atom cells  and zone center Brillouin zone sampling.
Similar calculations for Ce-doped systems were, however, found to yield erroneous results for the excited state.
In these calculations, the $4f-5d$ gap strongly (and nonphysically) decreases with increasing mixing parameter, leading to a level crossover in the \Ceneu{} configuration already for mixing parameters below 0.32.
We attribute this behavior to the inability of the PBE0 functional (as well as other common hybrid functionals) to account for differences in the screening of $4f$ and $5d$ states, as already noted in earlier studies on LaBr3$_3$ \cite{AbeSadErh12, ErhSadSch15}.
In all calculations involving Ce reported below, we therefore used exclusively the DFT+$U$ method.
Where necessary (\autoref{sect:results-ionization}), band gap errors were compensated by using the PBE$\rightarrow$PBE0 band edge shifts reported above as described in Ref.~\citenum{LinErhWah15}.

\section{Results}

\subsection{Ground and excited state landscapes}
\label{sect:ground-and-excited}

YAG crystallizes in space group I$a\bar{3}d$ (ITC number 230) with atoms on Wyckoff sites $24c$ (Y), $16a$/$24d$ (Al), and $96h$ (O).
The calculated lattice parameter is \unit[12.108]{\AA} to be compared with an experimental value of \unit[12.01$\pm$0.02]{\AA} \cite{YodKei51}.

\begin{figure}
    \centering
    \includegraphics{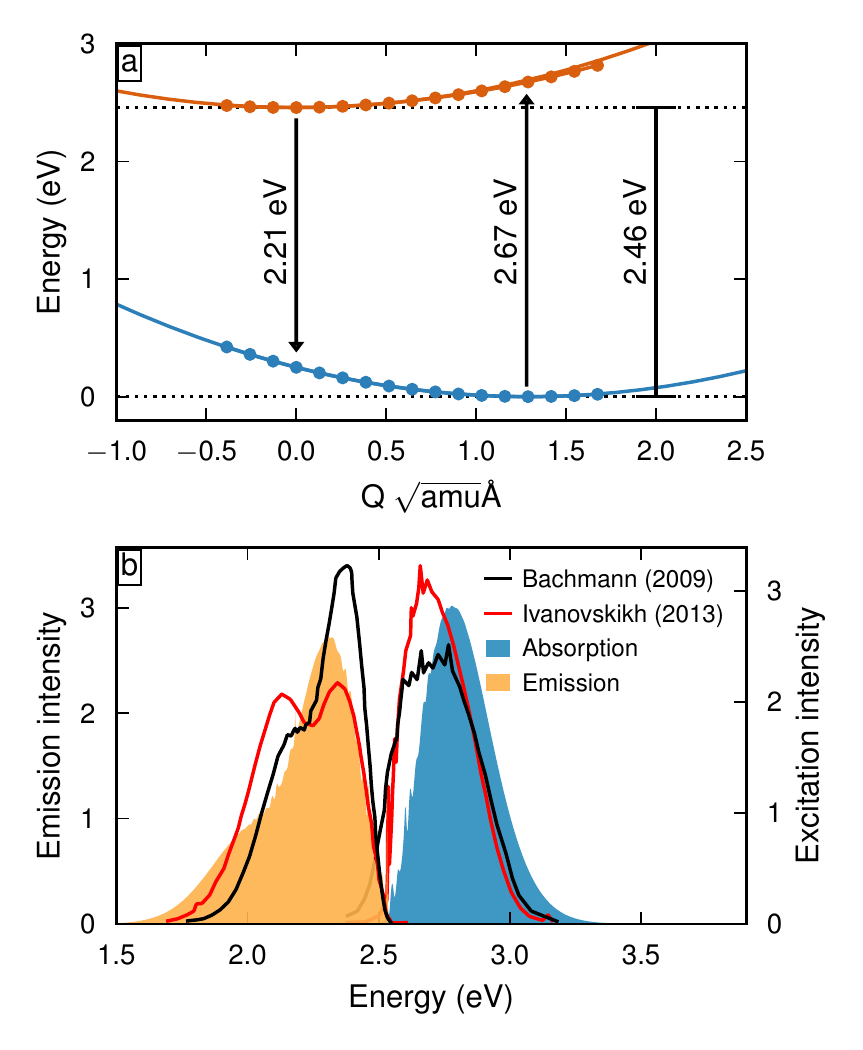}
    \caption{
        a) 1D configuration coordinate diagram for the $4f-5d$ transition.
        Solid lines are quadratic fits to the potential energies.
        b) Normalized optical lineshapes as computed with the generating function method.
        Black lines are experimental measurements from Ref.~\citenum{BacRonMei09} performed at \unit[5]{K} for the emission spectrum and at \unit[300]{K} for the absorption spectrum. Red lines are low temperature optical spectra from Ref.~\citenum{IvaOgiZyc2013}.
        The computed optical lineshapes have been shifted to the experimental zero-phonon lines of Ref.~\citenum{BacRonMei09} to simplify comparison.
    }
    \label{fig:configuration-coordination-diagram}
\end{figure}

In the ground state (Ce:$4f^15d^0$), the substitution of Ce on an Y site causes an outward relaxation of the nearest oxygen sites by on average \unit[0.07]{\AA}.
If the $4f$ electron is promoted to a $5d$ state, it becomes more delocalized, which effectively reduces the size of the Ce ion and causes an inward relaxation of the oxygen neighbors by \unit[0.06]{\AA} on average, such that the local geometry is almost identical to the Ce-free lattice.
The \gls{cc} diagram is then obtained by linear interpolation between the fully relaxed ground and excited state configurations (\autoref{fig:configuration-coordination-diagram}).
From the total energies we obtain an absorption energy of \unit[2.67]{eV} and an emission energy of \unit[2.21]{eV} in very good agreement with experimental values of 2.70 and \unit[2.31]{eV} \cite{BacRonMei09}.
The Stokes shifts are 0.25 and \unit[0.22]{eV} for emission and absorption, respectively, again in good agreement with the experimental value of \unit[0.30]{eV} \cite{BacRonMei09}.
Finally, our results are in agreement with earlier calculations of the absorption and emission energies with small quantitative differences due to different computational parameters \cite{JiaPonMig17}.

We note that in this case the energy differences between the $4f$ and $5d$ Kohn-Sham levels do not even semi-quantitatively correspond to the transition energies obtained from the total energies (see Fig.~S1).
This highlights the fact that while the Kohn-Sham levels can often be helpful starting points for quasi-particle descriptions, in general they cannot be interpreted as quasi-particle energies themselves.

A rough estimate for the crossing point between ground and excited state landscapes can be obtained by fitting a quadratic polynomial to the energies in the vicinity of the respective equilibrium geometries.
This yields a barrier of \unit[3.75]{eV}.
While this is a rough approximation of the upper limit, this very high energy barrier nonetheless strongly suggests that \emph{thermally activated} crossing of landscapes is unlikely to be an important factor, in agreement with earlier assessments \cite{Web73}.
The estimated energy barrier is an upper limit, and it is likely that the energy barrier will be smaller in more sophisticated models that go beyond the one-dimensional configurational coordinate model.

\subsection{Vibrational coupling to the \texorpdfstring{$4f-5d$}{4f-5d} transition}

The vibrational broadening of absorption and emission spectra is an important feature for the functionality of any material used for photon down-conversion.
It also provides important information concerning the energy landscape connecting the ground and excited state geometries.
As a result, a lot of spectroscopic data is available for \ceyag{}, which provides an opportunity to validate our calculations in more detail and to gain further insight into the energy landscape.

Our calculated spectra are in good overall agreement with experimental data \cite{BacRonMei09, IvaOgiZyc2013} (\autoref{fig:configuration-coordination-diagram}b) and the Stokes shifts computed from the spectra as the difference between the absorption maximum at \unit[2.71]{eV} and the emission intensity maximum at \unit[2.24]{eV} are close to the values obtained from the \gls{cc} diagram (\autoref{fig:configuration-coordination-diagram}a).

\begin{figure}
    \centering
    \includegraphics[width=\linewidth]{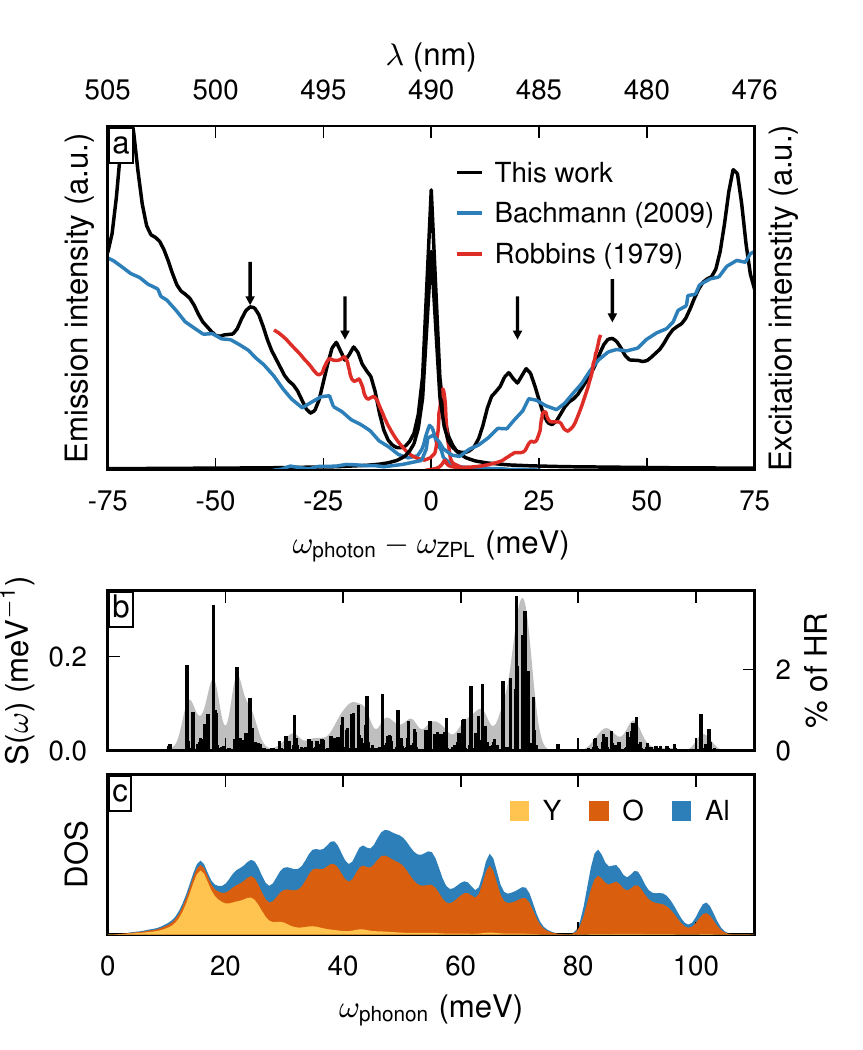}
    \caption{
        a) Fine structure of the lineshape around the \gls{zpl}.
        Experimental data is taken from Ref.~\citenum{BacRonMei09} (Bachmann (2009)) and Ref.~\citenum{Rob79} (Robbins (1979)).
        Experimental intensities were scaled to match the intensity of the computed spectra.
        The \gls{zpl} in the computed results have been shifted to the experimental value of Ref.~\citenum{BacRonMei09}.
        b) Electron-phonon spectral function (shaded) and the corresponding partial \gls{hr} factors.
        c) Phonon density of states decomposed into contributions from the different atomic species.
    }
    \label{fig:fine-structure}
\end{figure}

The fine structure of the spectra in the vicinity of the \gls{zpl} provides important information concerning the phonon modes that couple to the electronic transitions as well as the \gls{cc} diagram such as the total \gls{hr} factor.
Our calculations successfully reproduce the experimentally observed features in the fine structure of the optical spectra \cite{Rob79, BacRonMei09} also in this regard (\autoref{fig:fine-structure}a), including the total \gls{hr} factor, for which we obtain 5.5 using the normal modes for the ground state, to be compared with an experimental value of 6 \cite{BacRonMei09}.

Based on the experimental measurements \cite{Rob79, BacRonMei09}, it has been proposed that the fine structure of the optical spectra originates from a single mode around \unit[25]{meV} (or $\unit[200]{cm^{-1}}$) with higher energy features being replicas of this mode.
The presence of such a, presumably localized, mode would then also largely coincide with the \gls{cc}.
We can query the validity of this interpretation by analyzing the spectral function $S(\omega)$, which underlies the lineshape calculation (\autoref{sect:methodology}) and whose structure is reflected in the optical spectra.
This analysis reveals that the fine structure of the optical spectra originates from coupling to a large number of modes (\autoref{fig:fine-structure}b) in contrast to the earlier interpretations alluded to above \cite{Rob79, BacRonMei09}.

The main features are two distinct bands at approximately \unit[20]{meV} and \unit[70]{meV} and a less pronounced band at around \unit[40]{meV}.
The \unit[20]{meV} band is associated with \emph{dispersive} yttrium dominated modes while the \unit[70]{meV} band is due to \emph{dispersive} oxygen dominated modes, including motion of the nearest neighbor oxygen atoms of Ce.
At \unit[20]{meV}, the phonon density of states is dominated by modes associated with Y motion (\autoref{fig:fine-structure}c).
Importantly, there are, however, no significant contributions from \emph{localized} (defect) modes and there is no mode that contributes more than 4\% to the total \gls{hr} factor.
The calculations thus demonstrate that the fine structure of the optical spectra can be entirely explained by delocalized modes and that there is no distinct localized mode that can account for the relaxation along the \gls{cc}.

\subsection{Ionization of the \texorpdfstring{$5d$}{5d} state}
\label{sect:results-ionization}

Having addressed the character of ground and excited landscapes and its approximate connection to thermally activated landscape crossover (mechanism \emph{i}), we now address the thermal ionization model (mechanism \emph{iii}).
According to the latter, transitions from the excited $5d$ state to the conduction band manifold would result in luminescence quenching if the electron is subsequently captured at another site.

\begin{figure}
    \centering
    \includegraphics[scale=0.9]{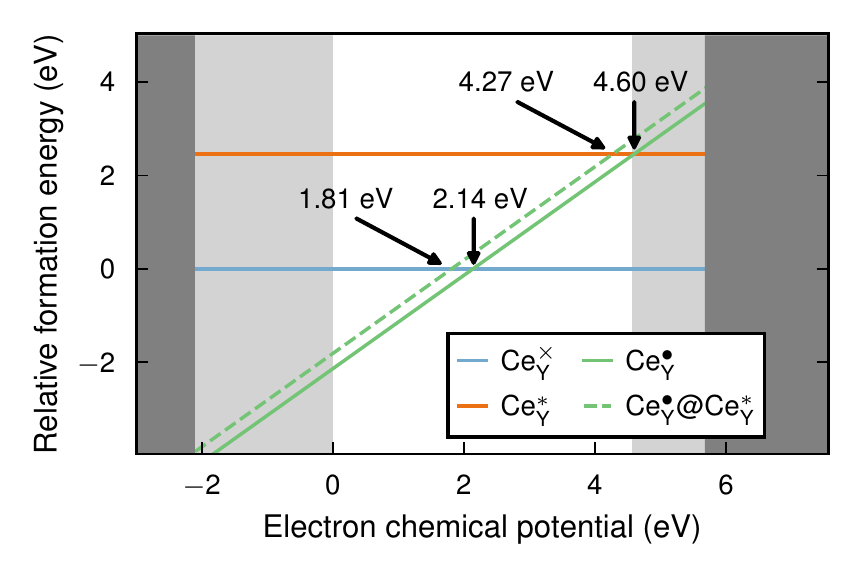}
    \caption{
        Relative formation energy of the Ce$_\text{Y}$ substitutional defect as a function of electron chemical potential.
        The PBE0 band edge shifts have been superimposed on the PBE values.
    }
    \label{fig:ef-ce}
\end{figure}

To evaluate the activation energy for ionization of the $5d$ state, we require its position with respect to the conduction band edge.
As noted above (\autoref{sect:ground-and-excited}), the Kohn-Sham eigenvalues do, however, not provide sensible guidance in the present case.
We therefore again resort to total energy differences instead and more precisely the relative defect formation energies (\autoref{fig:ef-ce}) as the ionization barrier corresponds to the \gls{ctl} from \Cepos{} to \Ceex{}.
Taking into account the band edge shift from PBE to PBE0, this yields a value of \unit[1.42]{eV}, when the \Cepos{} state is fixed at the \Ceex{} ionic configuration, corresponding to a vertical transition, or photoionization.
If the relaxation of \Cepos{} is taken into account, \gls{ctl} moves \unit[0.33]{eV} closer to the \gls{cbm} resulting in a transition energy of \unit[1.08]{eV}.
These values are in good agreement with estimations of the $5d$-\gls{cbm} distance of 0.76 to \unit[1.24]{eV} \cite{UedDorBos15, HamGayPog89}.
The ionization energy is likely to be reduced at higher temperatures due to renormalization of the conduction band edge.
This would place our prediction closer to the lower experimental value.

\begin{figure*}[!hbpt]
    \centering
    \includegraphics{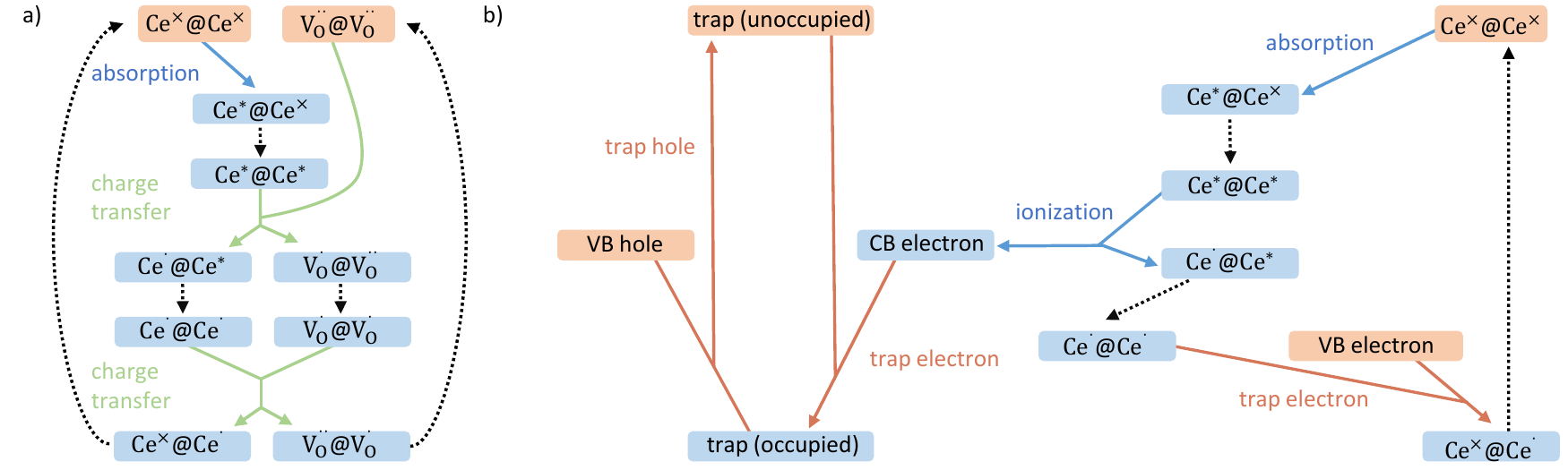}
    \caption{
        Illustration of reaction paths.
        a) Thermal quenching via charge transfer to an oxygen vacancy.
        b) Thermal quenching via electron transfer from Ce-$5d$ state to conduction band with subsequent trapping at a killer center.
    }
    \label{fig:reaction-paths}
\end{figure*}

\subsection{Charge transfer to an oxygen vacancy}
\label{sect:results-charge-transfer}

In many oxides, including YAG, \cite{CheLuXu08, ZheBinJie12} oxygen vacancies (\VO{}) are among the most important intrinsic defects \cite{Erh12, DemHolOha15, LinLinErh18, NikLagVed08}.
In wide band gap oxides, oxygen vacancies commonly feature deep \glspl{ctl} as well as electronic levels and exhibit substantial structural relaxation between different charge states \cite{LinLinErh18}.
These defects could thus play important roles in non-radiative recombination processes.
Here, we therefore consider in detail the charge transfer reaction from Ce to \VO{} (\autoref{fig:reaction-paths}a) and assess the ability of oxygen vacancies to act as electron traps in the ionization mechanism (\autoref{fig:reaction-paths}b).
We emphasize that even if oxygen vacancies are predominantly neutral for Fermi levels in the upper half of the band gap (the expected outcome after synthesis), positively charged oxygen vacancies are still the most probable donor defects by comparison with other candidates such as interstitials and antisites (also see Fig.~S3).
They are thus required to maintain charge neutrality (balancing free electrons at the conduction band edge) and electrically active.
In the following we therefore consider oxygen vacancies in charge states +2, +1 and 0.
(Negatively charged oxygen vacancies dissociate into a neutral vacancy and a free electron, see Fig.~S9).

The localized defect level associated with \VOneu{} lies inside the band gap, the associated charge density is localized in real space and has predominantly $s$-character, while the +2/0 \gls{ctl} resides \unit[1.79]{eV} below the \gls{cbm} at the PBE level and \unit[3.00]{eV} at the PBE0 level (see \gls{si} for the formation energy as a function of the electron chemical potential).
These observations are consistent with the oxygen vacancy inducing a deep defect state as anticipated \cite{LinLinErh18}.
We note that the position of the +2/0 \gls{ctl} is too deep to be identified as the electron traps experimentally found between \unit[0.86]{eV} and \unit[1.52]{eV} \cite{UedDorBos15}.
In \gls{LuAG} it was found that antisites can act as shallow electron traps \cite{NikVedFas07}.
In YAG, however, the antisites do not exhibit charge transition levels in the band gap (see \gls{si}).
Even so, the $\text{Al}_{\text{Y}}^{'}$ defect exhibits a localized Kohn-Sham state \unit[0.3]{eV} below the conduction band edge which can indicate that it can act as a local electron acceptor.

\begin{figure}
    \centering
    \includegraphics{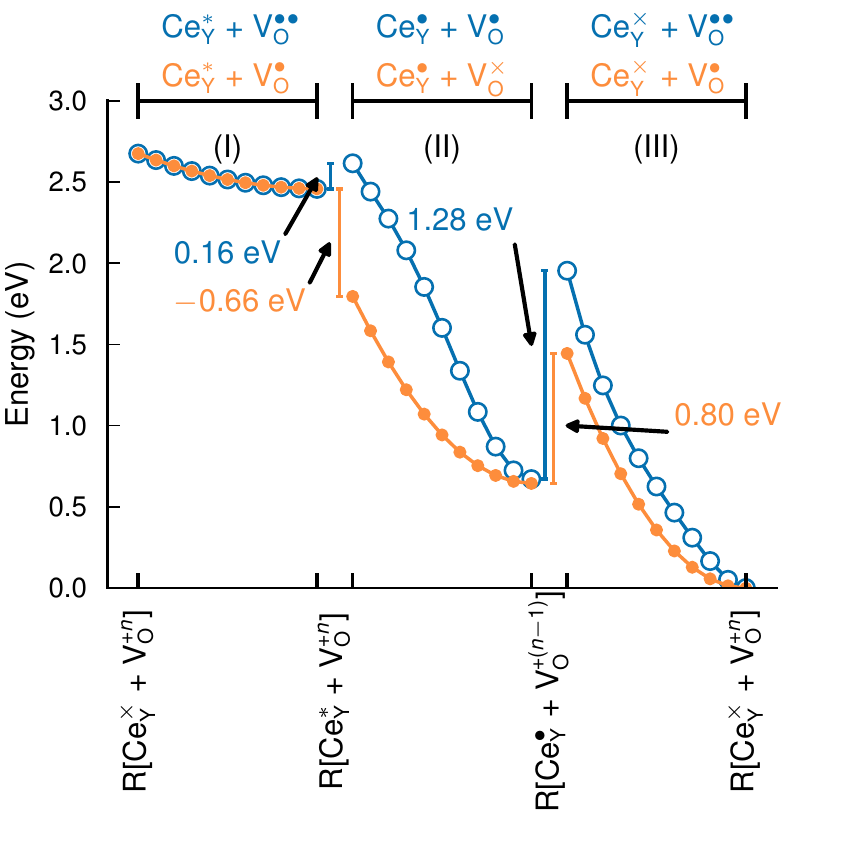}
    \caption{
        Thermal deactivation via oxygen vacancies.
        Points represent the sum of the total energy of Ce and V$_\text{O}$ defects for the same total charge states.
         The configurations on the abscissa refer to the equilibrium configuration of the respective combination of charge states.
    }
    \label{fig:energy-landscape}
\end{figure}

As a result of its electronic structure, the oxygen vacancy can participate in a charge transfer mechanism with Ce (\autoref{fig:energy-landscape}).
Immediately after absorption at a Ce site, the system undergoes a fast relaxation to the equilibrium geometry \Ceex{} dissipating \unit[0.22]{eV} in the process (\autoref{fig:configuration-coordination-diagram}a and relaxation along segment I in \autoref{fig:energy-landscape}).

At this point, the excitation can migrate through the system by resonant transfer between Ce sites.
This process is practical if the absorption and emission spectra overlap \cite{BlaGra94}, which is increasingly the case as the temperature goes up \cite{BacRonMei09}.
Alternatively, transfer can occur by non-zero exchange interaction.
This requires Ce sites to be sufficiently close to each other, which is the case if Ce concentrations reach the percent range.
Resonant transfer can proceed until emission occurs or until the excitation reaches a Ce atom that is in spatial proximity to either a singly (\VOpos{}) or doubly charged oxygen vacancy (\VOdpos{}).
In practice, the Ce concentration can be assumed to be much larger than the oxygen vacancy concentration, making it plausible that a large fraction of oxygen vacancies have a Ce atom in their proximity (right-most point of segment I in \autoref{fig:energy-landscape}).

After reaching a \Ceex{}--\VOpos{} or \Ceex{}--\VOdpos{} configuration, the electron in the \Ceex{} state can be transferred from the Ce atom to the oxygen vacancy (transition from segment I to segment II in \autoref{fig:energy-landscape}), leading to a charged Ce species (\Cepos{}) and a neutral (\VOneu{}) or singly charged vacancy (\VOpos{}).
The former transition leads to a reduction in the total energy by \unit[0.66]{eV} whereas the latter is associated with a small energy increase of \unit[0.16]{eV}.

The subsequent relaxation decreases the energy by \unit[1.15]{eV} (\Cepos{}--\VOneu{}) and \unit[1.94]{eV} (\Ceneu{}--\VOpos{}), respectively, which needs to be dissipated via coupling to lattice vibrations, leading to local heating (segment II in \autoref{fig:energy-landscape}).
The majority of this energy gain arises from the relaxation of the oxygen vacancy, while only \unit[0.33]{eV} are associated with the relaxation of the Ce site.
The large relaxation energy implies that the system is effectively trapped in a low energy state.
Already at this point, luminescence quenching has been achieved irreversibly.
The lowest energy pathway to escape from this configuration leads back to Ce in its ground state configuration \Ceneu{} in combination with either a singly or doubly charge oxygen vacancy with energy barriers of 0.80 and \unit[1.28]{eV}, respectively (transition from segment I to segment II in \autoref{fig:energy-landscape}).

\section{Discussion}

In the preceding sections, using first-principles calculations we obtained the limiting energy barriers for different recombination pathways.
Specifically, we obtained an approximate upper barrier of \unit[3.75]{eV} for the thermally activated landscape crossover (mechanism \emph{i}; \autoref{sect:ground-and-excited}) and a barrier of \unit[1.08]{eV} for the thermal ionization from occupied Ce-$5d$ states (mechanism \emph{iii}; \autoref{sect:results-ionization}).
In addition, we established a specific non-radiative recombination pathway in conjunction with resonant excitation transfer (mechanism \emph{ii}) that involves oxygen vacancies (\autoref{sect:results-charge-transfer}).
In the latter case, we found the maximum barriers along the reaction pathway to be \unit[0.80]{eV} for charge transfer via \VOpos{} and \unit[1.28]{eV} for charge transfer via \VOdpos{}.
The charge transfer mechanism via oxygen vacancies thus yields by far the lowest activation barrier among the mechanisms considered here (\autoref{fig:mechanisms}).

The energy barriers are on their own insufficient to quantitatively predict the rates for these recombination mechanisms.
They are, however, usually the most important parameter as for most mechanisms transition rates are exponentially dependent on the activation energy \cite{LinXiLin12}.
The charge transfer mechanism proposed here should thus be an important contribution to the non-radiative recombination in Ce-doped YAG.

Oxygen vacancies in YAG exhibit very strong and localized electron-phonon coupling.
The latter can be attributed to pronounced charge localization and the freedom for the atoms surrounding the vacancy to relax.
This allows oxygen vacancies to dissipate large amounts of energy locally while acting as a local acceptor relative to \Ceneu{}.
The limiting step in this mechanism is the transition from \Cepos{}--\VOneu{} to \Ceneu{}--\VOpos{}, at which point Ce is oxidized and therefore optically inactive.

In support of the charge transfer mechanism, it has been shown by X-ray absorption near edge structure that \Cepos{} can coexist with \Ceneu{} in \ceyag{} \cite{DanTesHom18} and that presence of electron acceptors in the vicinity of Ce can oxidize the Ce atom in related materials \cite{UedKatAsa17}.
One might suspect transition metal impurities to participate in a similar reaction.
The screening from the transition metal valence states as well as the much more restrictive geometry make it, however, unlikely that they dissipate similarly larger quantities of energy via a charge transfer mechanism.

The involvement of oxygen vacancies in the recombination process in Ce doped oxide phosphors has been considered before.
In Ref.~\citenum{JiaMigMas18}, the authors concluded that landscape crossover and $5d$ ionization could not satisfactorily explain the luminescence quenching in Lu$_2$SiO$_5$:Ce.
A hole autoionization process was therefore proposed in which \VOneu{} reduced the Ce atom leading to a $4f^15f^1$ electronic configuration .
While this model is conceptually different from the idea of oxygen vacancies as local energy dissipators presented here, it shares some features.
The hole autoionization model of Ref.~\citenum{JiaMigMas18} involves a reduction from \Ceneu{} to $\text{Ce}^{'}_\mathrm{Y}$ and an oxidation of the oxygen vacancy, while in the model proposed here the charge transfer in the first step of the reaction occurs in the opposite direction.

Finally, we note that the mechanism proposed in this study relies on general characteristics of oxygen vacancies in wide-gap oxides, most importantly their deep character, which entails strong relaxation between charge states and localized charges \cite{LinLinErh18}.
One can therefore anticipate it to be applicable also in other oxide based phosphors.
Moreover, the oxygen vacancy concentration can in principle be  controlled via the chemical environment during growth and annealing.
This knob could be used to further assess the role of oxygen vacancies in luminescence quenching.

\section{Conclusions}
\label{sect:conclusions}

To summarize, we have provided energy estimates for the most plausible luminescence quenching mechanisms in \ceyag{} by first-principles calculations.
The lowest energy pathway is obtained for a thermally activated concentration quenching mechanism that involves charge transfer between Ce atom and oxygen vacancies.
In this pathway, the limiting energy barrier is comparable to the thermal ionization energy.
As part of this investigation, we also analyzed the vibrational broadening and fine structure of the $4f-5d$ transition on Ce.
This analysis revealed that the fine structure of the phonon sidebands does not arise from a specific localized (defect) mode but is rather the result of a combination of many delocalized modes.

\section*{Acknowledgments}

We thank Maths Karlsson, Yuan-Chih Lin and Magnus Engholm for fruitful discussions.
Funding from the Knut and Alice Wallenberg Foundation (2014.0226) as well as the Swedish Research Council (2018-06482) are gratefully acknowledged.
The computations were enabled by resources provided by the Swedish National Infrastructure for Computing (SNIC) at NSC, C3SE and PDC partially funded by the Swedish Research Council through grant agreement no. 2018-05973.
Part of this work was performed under the auspices of the U.S. Department of Energy by Lawrence Livermore National Laboratory under Contract DE-AC52-07NA27344 with support from the National Nuclear Security Administration Office of Nonproliferation Research and Development (NA-22).

\end{document}